# Determination of Complex Refractive Index using Maximum Deviation Angle through Prism

Shosuke Sasaki

Center for Advanced High Magnetic Field Science, Graduate School of Science, Osaka University, Toyonaka, Osaka 560-0043, Japan

E-mail: sasaki@mag.ahmf.sci.osaka-u.ac.jp

Light-absorbing materials are widely used, and their optical properties are an important factor. Snell's law does not hold in materials that partially absorb light. Hence, the optical path in refraction is calculated from Maxwell's law. We used it to obtain the deviation angle when a light passes through a prism made of light absorbing material. As a result, the deviation angle has a local maximum point. The deviation angle near the local maximum is sensitive to the real and imaginary parts of the refractive index. The local maximum deviation angle and its incident angle are used to determine the complex index of refraction. This measurement has the same advantages as measuring the minimum deviation angle of a transparent prism. That is, when the optical bench is slightly rotated and the measurement light is observed, the moving direction of the light-spot is reversed at the extreme value. The detection is easy. Then, it is necessary to determine the complex refractive index from the measured local maximum deviation angle and its incident angle. These two angles are plotted parametrically by varying the real part of the refractive index under a fixed imaginary part. The similar curves are drawn under the fixed real parts. Each curve has a fixed value for the real or imaginary parts of the refractive index. Drawings of these many curves are made prior to measurement. Then we select the four curves that are closest to the local maximum deviation angle and its incident angle. Four fixed values attached to each of the four curves determine the complex index of refraction. Various liquids can be put in a prism container made of parallel plate glass and their local maximum deviation angles are measured. The method in this paper can easily determine the complex refractive index and can be used for material identification.

## 1. Introduction

With various substances, it is important to determine their optical properties. One of the representative quantities is the refractive index. Since the refractive index of light absorbing media is a complex number $n = n^{\text{Re}} + i\, n^{\text{Im}}$, Snell's law does not hold. In the references [1]-[3], the refraction phenomena were studied, but were not clarified in a light-absorbing medium. The previous paper [4] determined the optical path through a partially light-absorbing prism via the exact solution of Maxwell's law. Several studies [5]-[8] investigated the complex refractive index, but did not determine the optical path through a partially absorbing prism.

The previous paper [4] derived the complex index of refraction in two ways. The first method measures two angles. One is the exit angle X through the prism, when the incident light is perpendicular to the entrance surface. The second angle is the incident angle Y when the light exits normal to the exit surface. These two angles X and Y are different from each other. This is because the principle of optical path reversal does not hold due to the absorption of light. Two angles X and Y determine the complex index of refraction. Method 2 uses two prisms made of the same material with different apex angles. The second prism complicates the measurement. Therefore, we need to find an easier way.

The purpose of this paper is to find a simple method for determining the complex index of refraction. There is a local maximum of the deviation angle, when light passes through a prism made of partially absorbing maretial. We investigate how the deviation angle depends on the complex value of the refractive index. As the result of the investigation, the deviation angle varies sensitively in the vicinity of the local maximum, when the real and imaginary parts of the refractive index change. This property is suitable for determining the refractive index.

The local maximum deviation angle $\delta_{\text{max}}$ can be measured by a method similar to the conventional determination of the minimum deviation angle. It is difficult to obtain analytically the complex refractive index $n^{\text{Re}} + i\, n^{\text{Im}}$ from the measured local maximum deviation angle $\delta_{\text{max}}$ and its incident angle $\varphi_{\text{m}}$. However, it is possible to obtain the numerical values. We make the easiest way to do it. The complex index of refraction can be determined by using two sets of many curves. The horizontal axis of the figure is the incident angle $\varphi$ and the vertical axis is the deviation angle $\delta$. The points $(\varphi_{\text{m}}, \delta_{\text{max}})$ are plotted parametrically on the figure. The first series of curves are drawn with fixed values of the real part $n^{\text{Re}}$ by varing the imaginary part $n^{\text{Im}}$. Similarly, a second set of curves is drawn by gradually varing of $n^{\text{Re}}$ under fixed values of $n^{\text{Im}}$. We obtain a number of curves that intersect each other.

With the above preparations, the local maximum value of the deviation angle $\delta_{\text{max}}$ and its incident angle $\varphi_{\text{m}}$ are measured for a substance that partially absorbs light. We select the four closest curves that enclose the measured point $(\varphi_{\text{m}}, \delta_{\text{max}})$. Each curve has one fixed value for either real or imaginary part of the refractive index. The four values are taken as $n_1^{\text{Re}}, n_2^{\text{Re}}, n_1^{\text{Im}}$ and $n_2^{\text{Im}}$. Consequently, we get the inequalities $n_1^{\text{Re}} \leq n_{\text{m}}^{\text{Re}} \leq n_2^{\text{Re}}$ and $n_1^{\text{Im}} \leq n_{\text{m}}^{\text{Im}} \leq n_2^{\text{Im}}$ where $(n_{\text{m}}^{\text{Re}}, n_{\text{m}}^{\text{Im}})$ is the refractive index of the measured substance. Interpolating from the four values $n_1^{\text{Re}}, n_2^{\text{Re}}, n_1^{\text{Im}}, n_2^{\text{Im}}$ gives the complex refractive index $n_{\text{m}}^{\text{Re}} + i\, n_{\text{m}}^{\text{Im}}$ of the substance.

This method allows the use of conventional equipment for measuring the minimum deviation angle, and therefore, easily determine the complex refractive index of various substances. Especially for liquids, the complex refractive index can be measured by putting various liquids in a prism-shaped container made of parallel plate glass. Strongly absorbing materials transmit very little light, except for short distances through the prism. In such cases, the apex angle of the prism can be reduced for measurement.

## 2. Deviation angle of exiting light through prism

The previous paper [4] derived the optical path through a prism made of an absorbing medium. The method is complicated. This section derives a new method for determining the deviation angle through a prism. The method is simpler than the previous paper [4]. Figure 1 shows a schematic diagram of light passing through a prism with apex angle $\alpha$. The $x$-axis is to the right, the $y$-axis is from the back of the page to the front, and the $z$-axis is to the bottom.

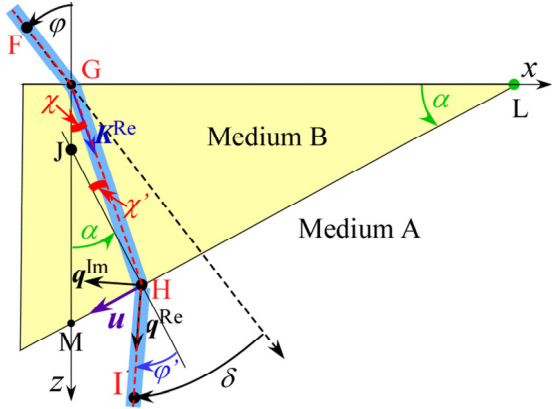

Fig.1 Optical path is FGHI. The angle $\alpha$ is the apex angle of the prism. Angle $\delta$ represents the deviation angle through the prism. $K$ and $q$ represent the wave vectors inside the prism and after exiting the prism. The two vectors $q^{Re}$ and $q^{Im}$ indicate the real and imaginary parts of $q$. The angles $\chi'$ ($\angle JHG$) and $\varphi'$ express the incident and refractive angles at the exit surface LHM.

Let $t$, $r$ and $\varphi$ be the time, position vector, and incident angle on the entrance surface of the prism, respectively. Let $\varepsilon_0$, $\mu_0$ and $c_0$ be the permittivity, magnetic permeability, and light velocity in vacuum, respectively. To simplify the expressions, we introduce two quantities $U$ and $V$ as

$$U = \left(\hat{\varepsilon}_B^{Re}\hat{\mu}_B^{Re} - \hat{\varepsilon}_B^{Im}\hat{\mu}_B^{Im}\right)/(\hat{\varepsilon}_A\hat{\mu}_A) \quad (1a)$$

$$V = \left(\hat{\varepsilon}_B^{Im}\hat{\mu}_B^{Re} + \hat{\varepsilon}_B^{Re}\hat{\mu}_B^{Im}\right)/(\hat{\varepsilon}_A\hat{\mu}_A) \quad (1b)$$

$$\hat{\varepsilon}_B = \hat{\varepsilon}_B^{Re} + i\,\hat{\varepsilon}_B^{Im}, \quad \hat{\mu}_B = \hat{\mu}_B^{Re} + i\,\hat{\mu}_B^{Im} \quad (2)$$

where $\hat{\varepsilon}_A$ and $\hat{\mu}_A$ express the relative permittivity and relative permeability of medium A in Fig.1. Also, $\hat{\varepsilon}_B$ and $\hat{\mu}_B$ are the relative permittivity and relative permeability of medium B in Fig.1. The real parts and imaginary parts are expressed by $\hat{\varepsilon}_B^{Re}$, $\hat{\varepsilon}_B^{Im}$, $\hat{\mu}_B^{Re}$ and $\hat{\mu}_B^{Im}$ [9],[10]. Let $E, B$, $E'', B''$, $E'''$ and $B'''$ be the incident electromagnetic fields, the refraction fields, and the exiting electromagnetic fields. They are expressed by

$$E = \widehat{E}e^{i(k\cdot r-\omega t)}, \quad B = \frac{1}{\omega}k \times E \quad \text{for } z < 0 \quad (3a)$$

$$E'' = \widehat{E}''e^{i(K\cdot r-\omega t)}, \quad B'' = \frac{1}{\omega}K \times E''$$
$$\text{for inside of prism} \quad (3b)$$

$$E''' = \widehat{E}'''\, e^{i(q\cdot r-\omega t)}, \quad B''' = \frac{1}{\omega}q \times E'''$$
$$\text{for the light exiting from prism} \quad (3c)$$

where $i$ is the imaginary unit, $\omega$ is the angular frequency, and the wave vector $k$ is

$$k = \frac{\omega}{c_0}\sqrt{\hat{\varepsilon}_A\,\hat{\mu}_A}(\sin\varphi, 0, \cos\varphi), \quad (4a)$$

$$\text{where} \quad c_0 = \frac{1}{\sqrt{\varepsilon_0\mu_0}} \quad (4b)$$

The parameter $\zeta$ was introduced in the previous papers [4], [11], [12] as follows;

$$\zeta = \arctan\left(V/(U - \sin^2\varphi)\right) \quad (5)$$

New parameter $W$ is introduced for simplicity of calculations.

$$W = \{(U - \sin^2\varphi)^2 + V^2\}^{1/4} \quad (6)$$

Then, the parameter $\zeta$ is expressed by

$$\zeta = \arcsin\left(\frac{V}{\sqrt{(U-\sin^2\varphi)^2+V^2}}\right) = \arcsin\left(\frac{V}{W^2}\right) \quad (7)$$

Using $W$, the deviation angle can be derived more easily than the previous calculation [4]. The wave vector $K$ is obtained from the boundary condition at the entrance surface of prism as,

$$K = K^{Re} + i\,K^{Im} = \frac{\omega\sqrt{\hat{\varepsilon}_A\hat{\mu}_A}}{c_0}\left(\sin\varphi, 0, We^{i\zeta/2}\right) \quad (8a)$$

$$\mathbf{K}^{\text{Re}} = \frac{\omega\sqrt{\hat{\varepsilon}_A\hat{\mu}_A}}{c_0}\left(\sin\varphi,\ 0,\ W\cos(\zeta/2)\right) \quad (8b)$$

$$\mathbf{K}^{\text{Im}} = \frac{\omega\sqrt{\hat{\varepsilon}_A\hat{\mu}_A}}{c_0}\left(0, 0,\ W\sin(\zeta/2)\right) \quad (8c)$$

The refraction angle $\chi$ is angle $\angle$JGH which is derived from Eq.(8b) as follows;

$$\chi = \arctan\left[\frac{\sin\varphi}{W\cos(\zeta/2)}\right] \quad (9)$$

The refraction light exits through the exit boundary MHL. In the exit boundary, the incident angle is $\chi'$ and the refracted angle is $\varphi'$. As easily seen in Fig.1, angle $\chi'$, namely $\angle$JHG is equal to $\alpha - \chi$, where $\alpha$ is the apex angle of the prism;

$$\chi' = \alpha - \chi \quad (10)$$

Since the electromagnetic field (3c) satisfies the wave equation, $\omega$ and $\mathbf{q} = \mathbf{q}^{\text{Re}} + i\,\mathbf{q}^{\text{Im}}$ are related as follows:

$$\mathbf{q} \cdot \mathbf{q} = \varepsilon_0\mu_0\hat{\varepsilon}_A\,\hat{\mu}_A\omega^2 \quad (11a)$$

$$\mathbf{q}^{\text{Re}} \cdot \mathbf{q}^{\text{Re}} - \mathbf{q}^{\text{Im}} \cdot \mathbf{q}^{\text{Im}} = \varepsilon_0\mu_0\hat{\varepsilon}_A\,\hat{\mu}_A\omega^2 \quad (11b)$$

$$\mathbf{q}^{\text{Im}} \cdot \mathbf{q}^{\text{Re}} = 0 \quad (11c)$$

Let $\mathbf{u}$ be the unit vector parallel to the line LHM in Fig.1.

$$\mathbf{u} = (-\cos\alpha, 0, \sin\alpha) \quad (12)$$

On the boundary LHM, the phase $i(\mathbf{q}\cdot\mathbf{u} - \omega t)$ outside the prism is equal to the phase $i(\mathbf{K}\cdot\mathbf{u} - \omega t)$ inside the prism due to the boundary condition of the electric field. Therefore,

$$\mathbf{q} \cdot \mathbf{u} = \mathbf{K} \cdot \mathbf{u} \quad (13a)$$

$$q_y = K_y = 0 \quad (13b)$$

Substitution of Eq.(8a) into the right hand of Eq.(13a) yields

$$\mathbf{q}\cdot\mathbf{u} = \frac{\omega\sqrt{\hat{\varepsilon}_A\hat{\mu}_A}}{c_0}\left(-\cos\alpha\,\sin\varphi + \sin\alpha\,W\,e^{i\zeta/2}\right) \quad (14a)$$

$$\mathbf{q}^{\text{Re}}\cdot\mathbf{u} = q^{\text{Re}}\sin\varphi' = \frac{\omega\sqrt{\hat{\varepsilon}_A\hat{\mu}_A}}{c_0}(-\cos\alpha\,\sin\varphi + \sin\alpha\,W\cos(\zeta/2)) \quad (14b)$$

$$\mathbf{q}^{\text{Im}}\cdot\mathbf{u} = q^{\text{Im}}\cos\varphi' = \frac{\omega\sqrt{\hat{\varepsilon}_A\hat{\mu}_A}}{c_0}\sin\alpha\,W\sin(\zeta/2) \quad (14c)$$

Squares of both sides of Eqs.(14b) and (14c) give the following equations;

$$(q^{\text{Re}})^2 = \frac{\omega^2\hat{\varepsilon}_A\hat{\mu}_A}{c_0^2\sin^2\varphi'} \times (-\cos\alpha\,\sin\varphi + \sin\alpha\,W\cos(\zeta/2))^2 \quad (15a)$$

$$(q^{\text{Im}})^2 = \frac{\omega^2\hat{\varepsilon}_A\hat{\mu}_A}{c_0^2\cos^2\varphi'}(\sin\alpha\,W\sin(\zeta/2))^2 \quad (15b)$$

Substitution of (15a) and (15b) into (11b) yields

$$\frac{F}{(\sin\varphi')^2} - \frac{G}{(\cos\varphi')^2} = 1 \quad (16a)$$

$$F = (-\cos\alpha\,\sin\varphi + \sin\alpha\,W\cos(\zeta/2))^2 \quad (16b)$$

$$G = (\sin\alpha\,W\sin(\zeta/2))^2 \quad (16c)$$

Equation (16a) produces a quadratic equation of $(\sin\varphi')^2$ as follows;

$$(\sin\varphi')^4 - (\sin\varphi')^2 - F(\sin\varphi')^2 - G(\sin\varphi')^2 + F = 0 \quad (17)$$

Because $(\sin\varphi')^2$ is smaller than 1, the solution is given by

$$(\sin\varphi')^2 = \tfrac{1}{2}(F + G + 1) - \tfrac{1}{2}\sqrt{(F + G + 1)^2 - 4F} \quad (18)$$

The sign of $\sin\varphi'$ is the same as the sign of $\sin\chi'$ as shown in Fig.1. Therefore,

$$\sin\varphi' = \sqrt{\tfrac{1}{2}\left(F + G + 1 - \sqrt{(F + G + 1)^2 - 4F}\right)} \quad \text{for } \chi' \geq 0 \quad (19a)$$

$$\sin\varphi' = -\sqrt{\tfrac{1}{2}\left(F + G + 1 - \sqrt{(F + G + 1)^2 - 4F}\right)} \quad \text{for } \chi' < 0 \quad (19b)$$

$$\varphi' = \arcsin\left[\sqrt{\tfrac{1}{2}\left(F + G + 1 - \sqrt{(F + G + 1)^2 - 4F}\right)}\right] \quad \text{for } \chi' \geq 0 \quad (20a)$$

$$\varphi' = -\arcsin\left[\sqrt{\tfrac{1}{2}\left(F + G + 1 - \sqrt{(F + G + 1)^2 - 4F}\right)}\right] \quad \text{for } \chi' < 0 \quad (20b)$$

The deviation angle $\delta$ through the prism is derived from Fig.1 as follows;

$$\delta = \varphi + \varphi' - \alpha \quad (21)$$

This derivation of $\delta$ is simpler than the previous method [4]. The parameters $F$ and $G$ are equivalent to $f$ and $g$ in the previous paper, respectively. I next investigate how the deviation angle depends on the refractive index, especially its imaginary part.

## 3. Dependence of deviation angle on refractive index

The refractive indexes $n_A$ and $n_B$ of the substances A and B are defined by

$$n_A = \sqrt{\hat{\varepsilon}_A \hat{\mu}_A} \tag{22a}$$

$$n_B = \sqrt{\hat{\varepsilon}_B \hat{\mu}_B} \tag{22b}$$

The relative refractive index $n_{BA}$ between B and A is expressed by

$$n_{BA} = n_B/n_A = \sqrt{U + iV} \tag{23}$$

The four parameters $U, V, W$ and $\zeta$ are given by the relative refractive index $n_{BA} = n_{BA}^{Re} + i\, n_{BA}^{Im}$ and the incident angle $\varphi$ as follows;

$$U = \left(n_{BA}^{Re}\right)^2 - \left(n_{BA}^{Im}\right)^2 \tag{24a}$$

$$V = 2\, n_{BA}^{Re}\, n_{BA}^{Im} \tag{24b}$$

$$W = \{((n_{BA}^{Re})^2 - (n_{BA}^{Im})^2 - \sin^2\varphi)^2 + (2\, n_{BA}^{Re}\, n_{BA}^{Im})^2\}^{1/4} \tag{24c}$$

$$\zeta = \arcsin\left(\frac{V}{W^2}\right) \tag{24d}$$

The deviation angle $\delta$ versus incident angle $\varphi$ is calculated from Eqs.(16b), (16c), (20a), (20b) and (21). Figure 2 shows the twenty curves of $(\varphi, \delta)$ via a prism with apex angle $60°$. The red curves express for $n_{BA}^{Re} = 1.7$ and the black curves for $n_{BA}^{Re} = 1.4$. The imaginary part $n_{BA}^{Im}$ takes ten values $0.005 \sim 0.05$.

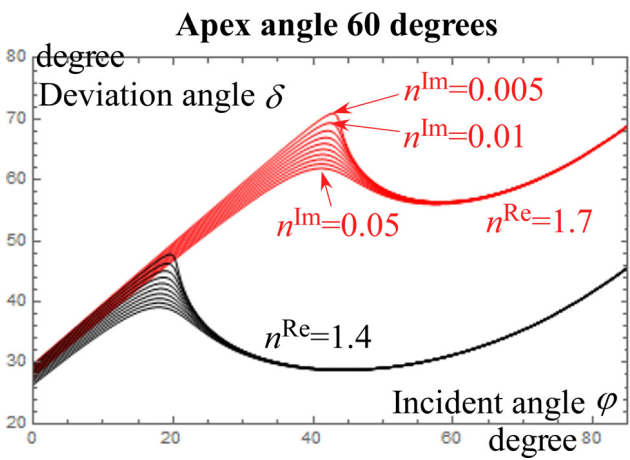

Fig.2: Deviation angle versus incident angle through a prism with apex angle $60°$. Red curve for $n_{BA}^{Re} = 1.7$. Black curves for $n_{BA}^{Re} = 1.4$. Both curves have ten values of $n_{BA}^{Im} = 0.005 \sim 0.05$ in 0.005 step.

As can be easily seen in Fig.2, the deviation angle $\delta$ varies sensitively with a change of $n_{BA}^{Im}$ near the local maximum point. Figure 3 shows 20 curves with $n_{BA}^{Im}$ that are two orders of magnitude smaller than Fig.2. Angle measurement accuracy is required, but it has the advantage of being able to measure the material-specific values $n_{BA}^{Re}$ and $n_{BA}^{Im}$.

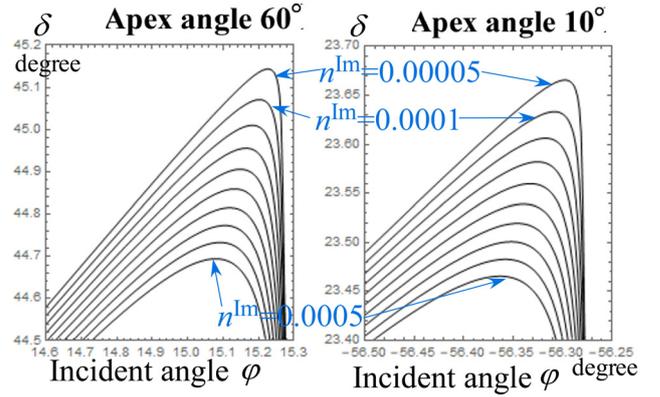

Fig.3: Left panel expresses 10 curves of $(\varphi, \delta)$ through a prism with apex angle $60°$. Right panel with apex angle $10°$. Both panels show the curves of $(\varphi, \delta)$ in the cases of $n_{BA}^{Re} = 1.333$ and $n_{BA}^{Im} = 0.00005 \sim 0.0005$.

Even if the absorption is large (the intensity of the transmitted light is weak), it can be handled by shortening the transmission distance. Specifically, it is preferable to reduce the apex angle of the prism.

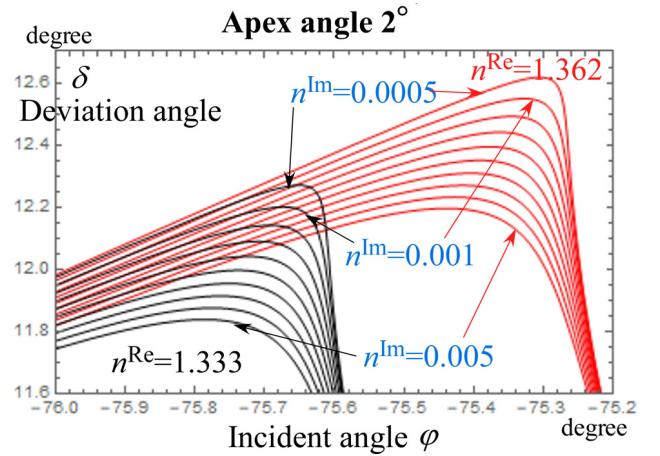

Fig.4: Deviation angle versus incident angle, through a prism with apex angle $2°$. $n_{AB}^{Re} = 1.362$ in red curves. $n_{AB}^{Re} = 1.333$ in black curves. $n_{BA}^{Im} = 0.0005 \sim 0.005$.

Figure 4 shows the curves of deviation angle depending on the incident angle for a prism with apex angle 2°. As examples, the real part of the refractive index is chosen to be the same as ethanol and water. (Of course, the deviation angle versus incident angle can be drawn for any complex refractive index.) The red curves are at $n_{BA}^{Re} = 1.362$, The black curves are at $n_{BA}^{Re} = 1.333$. Both curves have 10 imaginary parts of $n_{BA}^{Im} = 0.0005 \sim 0.005$. Prism-shaped containers made of plate glass are prepared which have different apex angles (60°, 10°, 2°, etc.). Various liquids are put into them and then, the complex refractive index can be measured in a wide range of light-absorption by the liquids. In the next section, we establish a method to determine the complex index of refraction from measurements of the local maximum deviation angle $\delta_{max}$ and its incident angle $\varphi_m$.

## 4. Determination of complex refractive index

The local maximum deviation angle $\delta_{max}$ and its incident angle $\varphi_m$ depend on the relative complex refractive index $n_{BA}^{Re} + i\, n_{BA}^{Im}$. The points $(\varphi_m, \delta_{max})$ are drawn by varying the two parameters $(n_{BA}^{Re}, n_{BA}^{Im})$.

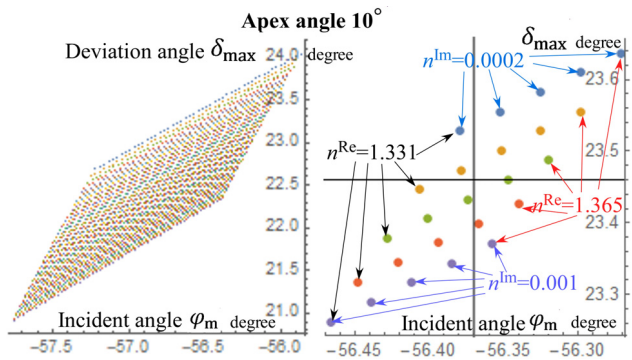

Fig.5 Points $(\varphi_m, \delta_{max})$ are shown for apex angle 10°.
Left: $n_{BA}^{Re} = 1.301 \sim 1.350$, $n_{BA}^{Im} = 0.0002 \sim 0.0100$
Right: $n_{BA}^{Re} = 1.331 \sim 1.335$, $n_{BA}^{Im} = 0.0002 \sim 0.0010$

The left panel of Fig.5 shows the points $(\varphi_m, \delta_{max})$ at the complex refractive indexes for 2500 cases:

$n_{BA}^{Re} = 1.3 + 0.001 \times m$ for $m = 1, 2, \cdots 50$ (25a)
$n_{BA}^{Im} = 0.0002 \times n$ for $n = 1, 2, \cdots 50$ (25b)

The right panel shows 25 points in Eqs.(25a) and (25b) with $m=1,2,\cdots 5$ and $n=1,2,\cdots 5$. The values of 25 points $(\varphi_m, \delta_{max})$ are listed in Table I. The inside of the table is divided into two parts. The upper part expresses the incident angle $\varphi_m$ and the lower part is the local maximum deviation angle $\delta_{max}$. The listed values $(\varphi_m, \delta_{max})$ are corresponding to 25 points in the right panel of Fig.5.

Table I: Local maximum deviation angle and its incident angle. Apex angle of prism is 10°. The top row is the values of real part $n_{BA}^{Re}$ and the leftmost column expresses the imaginary part $n_{BA}^{Im}$. The inside of the table is divided into two parts, the upper part is $\varphi_m$ and the lower part is $\delta_{max}$. Angle units are degrees.

|  | 1.331 | 1.332 | 1.333 | 1.334 | 1.335 |
|---|---|---|---|---|---|
| $n_{BA}^{Im} =$ | -56.380 | -56.352 | -56.325 | -56.298 | 56.271 |
| 0.0002 | 23.528 | 23.555 | 23.582 | 23.609 | 23.637 |
| $n_{BA}^{Im} =$ | -56.406 | -56.379 | -56.352 | -56.325 | -56.298 |
| 0.0004 | 23.446 | 23.474 | 23.501 | 23.528 | 23.555 |
| $n_{BA}^{Im} =$ | -56.429 | -56.401 | -56.374 | -56.347 | -56.320 |
| 0.0006 | 23.378 | 23.405 | 23.433 | 23.460 | 23.487 |
| $n_{BA}^{Im} =$ | -56.448 | -56.421 | -56.394 | -56.367 | -56.340 |
| 0.0008 | 23.317 | 23.345 | 23.372 | 23.399 | 23.427 |
| $n_{BA}^{Im} =$ | -56.466 | -56.439 | -56.412 | -56.385 | -56.358 |
| 0.001 | 23.261 | 23.289 | 23.316 | 23.344 | 23.371 |

**(Example of how to determine the complex index of refraction)**

As an example, let the measured value of the incident angle and the local maximum deviation angle be the following values;

$$(\varphi_m, \delta_{max}) = (-56.37°, 23.46°) \quad (26)$$

This point is indicated by a cross line in the right panel of Fig.5. We find the four closest points to the point of Eq.(26). The values of $n_{BA}^{Re}$ are 1.332 and 1.333, and the values of $n_{BA}^{Im}$ are 0.0004 and 0.0006. These four points are numbered as follows:

For $n_1^{Re} = 1.332$, $n_1^{Im} = 0.0004$,
$$(\varphi_1, \delta_1) = (-56.379°, 23.474°) \quad (27a)$$

For $n_2^{Re} = 1.333$, $n_2^{Im} = 0.0004$,
$$(\varphi_2, \delta_2) = (-56.352°, 23.501°) \quad (27b)$$

For $n_3^{Re} = 1.332$, $n_3^{Im} = 0.0006$,
$$(\varphi_3, \delta_3) = (-56.401°, 23.405°) \quad (27c)$$

For $n_4^{Re} = 1.333$, $n_4^{Im} = 0.0006$,
$$(\varphi_4, \delta_4) = (-56.374°, 23.433°) \quad (27d)$$

The difference between $(\varphi_m, \delta_{max})$ and $(\varphi_1, \delta_1)$ is represented by the partial derivatives as follows;

$$\varphi_m - \varphi_1 = \frac{\partial \varphi}{\partial n^{Re}} \Delta n^{Re} + \frac{\partial \varphi}{\partial n^{Im}} \Delta n^{Im} \quad (28a)$$

$$\delta_{max} - \delta_1 = \frac{\partial \delta}{\partial n^{Re}} \Delta n^{Re} + \frac{\partial \delta}{\partial n^{Im}} \Delta n^{Im} \quad (28b)$$

The partial derivatives are given from Eqs.(27a), (27b) and (27c)

$$\frac{\partial \varphi}{\partial n^{Re}} = \frac{0.027}{0.001} = 27, \quad \frac{\partial \varphi}{\partial n^{Im}} = \frac{-0.022}{0.0002} = -110 \quad (29a)$$

$$\frac{\partial \delta}{\partial n^{Re}} = \frac{0.027}{0.001} = 27, \quad \frac{\partial \delta}{\partial n^{Im}} = \frac{-0.069}{0.0002} = -345 \quad (29b)$$

Equations (28a) and (28b) yields,

$$\begin{pmatrix} \varphi_m - \varphi_1 \\ \delta_{max} - \delta_1 \end{pmatrix} = \begin{pmatrix} \frac{\partial \varphi}{\partial n^{Re}} & \frac{\partial \varphi}{\partial n^{Im}} \\ \frac{\partial \delta}{\partial n^{Re}} & \frac{\partial \delta}{\partial n^{Im}} \end{pmatrix} \begin{pmatrix} \Delta n^{Re} \\ \Delta n^{Im} \end{pmatrix} \quad (30)$$

Substitution of Eqs.(29a) and (29b) into Eq.(30) gives

$$\begin{pmatrix} 0.009 \\ -0.014 \end{pmatrix} = \begin{pmatrix} 27 & -110 \\ 27 & -345 \end{pmatrix} \begin{pmatrix} \Delta n^{Re} \\ \Delta n^{Im} \end{pmatrix} \quad (31)$$

Solution of Eq.(31) is

$$\begin{pmatrix} \Delta n^{Re} \\ \Delta n^{Im} \end{pmatrix} = \begin{pmatrix} 0.000732 \\ 0.0000979 \end{pmatrix} \quad (32)$$

Therefore, the refractive index is obtained as,

$$n^{Re} = n_1^{Re} + \Delta n^{Re} \approx 1.3327 \quad (33a)$$
$$n^{Im} = n_1^{Im} + \Delta n^{Im} \approx 0.000498 \quad (33b)$$

If the measurement accuracy is low, it is sufficient to simply search for a refractive index with a similar value of $(\varphi_m, \delta_{max})$ from the table. A more extensive table of the list $(n^{Re}, n^{Im}, \varphi_m, \delta_{max})$ is attached to this paper.

## 5. Conclusions

A new phenomenon has appeared in the refraction of light passing through a prism that partially absorbs it. The optical path through the prism was determined by solving Maxwell's equations. Then, the local maximum deviation angle $\delta_{max}$ appears. The angle is sensitive to changes in the refractive index. This paper showed how the local maximum deviation angle and its incident angle determine the complex refractive index $n_{BA}^{Re} + i\, n_{BA}^{Im}$. The method for determining the complex index of refraction used Table I, which is the relationship between $(n_{BA}^{Re}, n_{BA}^{Im})$ and $(\varphi_m, \delta_{max})$. To apply this method to a variety of materials, the list of $(n_{BA}^{Re}, n_{BA}^{Im}, \varphi_m, \delta_{max})$ needs to be extended to a wider range and finer steps for the complex refractive indices. For the apex angle of $10°$, the list is attached as a csv file to this paper. The range of $n_{BA}^{Re}$ is 1.2 to 1.8 in steps of 0.0005, and the range of $n_{BA}^{Im}$ is 0.0001 to 0.01 in steps of 0.0001. For other apex angles, please request the author to create and send.